\documentstyle[ltwol,psfig,axodraw]{article}

\def\Journal#1#2#3#4{{#1} {\bf #2}, #3 (#4)}

\def\NPB{{\em Nucl. Phys.} B}
\def\PLB{{\em Phys. Lett.}  B}

\def\PRD{{\em Phys. Rev.} D}

\def\EPJC{{\em Eur. Phys. J.} C}

\def\be{\begin{equation}}
\def\ee{\end{equation}}
\def\bea{\begin{eqnarray}}
\def\eea{\end{eqnarray}}

\def\r2{\sqrt{2}}
\def\gsim{~{\rlap{\lower 3.5pt\hbox{$\mathchar\sim$}}\raise 1pt\hbox{$>$}}\,}
\def\BSgamma{B\rightarrow X_s\gamma}
\def\BsBs{B_s^0\overline{B_s^0}}
\def\mqu#1{m_{u#1}}
\def\mqd#1{m_{d#1}}
\def\VT{V_{32}^*V_{33}}
\def\VU{V_{42}^*V_{43}}
\def\ZBS{(V^\dagger V)_{23}}

\begin{document}

\title{CONTRIBUTIONS OF VECTOR-LIKE QUARKS TO 
$\BSgamma$ AND $\BsBs$ MIXING}

\author{MAYUMI AOKI}

\address{Theory Group, KEK, Tsukuba, Ibaraki 305-0801, Japan \\
E-mail: mayumi.aoki@kek.jp}   

\author{MAKIKO NAGASHIMA}

\address{Graduate School of Humanities and Sciences, 
Ochanomizu University, Bunkyo-ku, Tokyo 112-8610, Japan \\
E-mail: g0070508@edu.cc.ocha.ac.jp}  

\author{NORIYUKI OSHIMO}

\address{Department of Physics, 
Ochanomizu University, Bunkyo-ku, Tokyo 112-8610, Japan \\
E-mail: oshimo@phys.ocha.ac.jp}  


\twocolumn[\maketitle\abstracts{
The radiative decay $\BSgamma$ and the mixing $\BsBs$ are 
discussed in the vector-like quark model which contains extra   
SU(2)-singlet quarks with the same electric charges as 
the up-type and down-type quarks.  
Constraints on the extended Cabibbo-Kobayashi-Maskawa matrix 
are obtained from the experimental results for the branching ratio
of $\BSgamma$ .   
Within these experimental bounds, any value of the branching ratio 
can be accounted for by the contributions from the vector-like quarks.   
In sizable ranges of the model parameters, the mixing parameter for $\BsBs$ 
is much different from the prediction of the standard model.  
}]

\section{Introduction}

     A number of reasons suggest that some extension 
of the standard model (SM) be necessary for describing physics 
around or above the electroweak energy scale.  
Since several phenomena involved in the $B$-meson system are 
sensitive to the extension of the SM, new physics may be unveiled by 
detailed examinations of these phenomena in the present or near-future 
experiments, such as B factories, BTeV, and LHCb.  
For instance, the radiative $B$-meson decay 
and $B^0\overline{B^0}$ mixing could receive non-trivial contributions 
from supersymmetry~\cite{oshimo,branco}.  
Systematic studies of various extensions of the SM should 
be performed in order to pinpoint a plausible model.  

     In this report we discuss the decay $\BSgamma$~\cite{aokibs} and 
the mixing $\BsBs$~\cite{aokibb}  
within the framework of the vector-like quark model (VQM) which is 
one of the minimal extensions of the SM.  
This model contains extra quarks with electric charges 2/3 and/or 
$-1/3$, whose left-handed components, as well as right-handed ones, 
are singlets under the SU(2) gauge transformation.  
Therefore, phenomena of flavor-changing neutral current (FCNC) 
are affected.   
It will be shown that the decay rate of $\BSgamma$ can deviate  
from the prediction of the SM.  
The experimental results for the decay rate available at present 
already impose certain constraints on the model.  
The amount of $\BsBs$ mixing can also 
be much different from the SM prediction.  

     In sect.~2 a brief summary of the VQM is presented.  
We consider the branching ratio of $\BSgamma$ in sect.~3, 
and the mixing parameter $x_s$ for $\BsBs$ in sect.~4.  
Discussions are given in sect.~5.  

\section{Generation Mixings}

     The VQM incorporates extra Dirac 
fermions which are transformed as $(3,1,2/3)$ and/or $(3,1,-1/3)$ under 
the SU(3)$\times$SU(2)$\times$U(1) gauge symmetry.  
The left-handed and right-handed components have the same properties.  
For definiteness, we assume that there exist one up-type 
and one down-type vector-like quarks.  
The masses of the quarks are generated through Yukawa couplings 
and bare mass terms.   
The whole mass terms are expressed by 4$\times$4 matrices, 
which are denoted by $M^u$ and $M^d$ respectively for up-type and 
down-type quarks.  
The mass eigenstates are obtained by diagonalizing 
the mass matrices as  
\bea
      A_L^{u\dagger} M^uA^u_R &=& {\rm diag}(\mqu1,\mqu2,\mqu3,\mqu4),   \\
      A_L^{d\dagger} M^dA^d_R &=& {\rm diag}(\mqd1,\mqd2,\mqd3,\mqd4),   
\eea
where $A^u_L$, $A^u_R$, $A^d_L$, and $A^d_R$ denote unitary matrices.  

     A large difference between the VQM and the SM resides in 
the Cabibbo-Kobayashi-Maskawa (CKM) matrix $V$ 
for the weak interactions.  
The CKM matrix of the VQM is enlarged to be a $4\times 4$ matrix, which is 
given by 
\be
      V_{ab} = \sum_{i=1}^3(A_L^{u\dagger})_{ai}(A_L^d)_{ib}.  
\ee
Since the left-handed components of the extra quarks are 
singlets under the SU(2) transformation, 
the matrix $V$ is not unitary:   
\be
  (V^\dagger V)_{ab}=\delta_{ab}-A^{d*}_{L4a}A^d_{L4b}.  
\ee
Consequently, the interactions of the quarks with the $W$ boson 
become qualitatively different from those in the SM.   
The $Z$ boson couples directly to the quarks with different flavors.  
The neutral Higgs boson also mediates 
flavor-changing interactions at the tree level.  

     The interactions mediated by the $W$, $Z$, and Higgs bosons 
give new contributions to processes of FCNC.  
However, taking into account the constraints on the VQM imposed from 
the presently available experiments,  
the effects by the $Z$ and Higgs bosons are found to be small.  
Our study is thus concentrated on the contributions coming from 
the $W$-mediated interactions.   
The interaction Lagrangian for the quarks with the $W$ and 
Goldstone bosons is given by 
\begin{eqnarray}
   {\cal L} &=& \frac{g}{\r2}\sum_{a,b=1}^4\overline{u^a}V_{ab}\gamma^\mu               
                   \frac{1-\gamma_5}{2}d^bW_\mu^\dagger   \nonumber  \\ 
 &+& \frac{g}{\r2}\sum_{a,b=1}^4\overline{u^a}V_{ab}\left\{\frac{m_{ua}}{M_W}  
                   \left(\frac{1-\gamma_5}{2}\right) \right.  \nonumber \\
  & & \left. - \frac{m_{db}}{M_W}\left(\frac{1+\gamma_5}{2}\right)\right\}
                d^bG^\dagger  \nonumber \\
         &+& {\rm h.c.}.    
\label{Wlagrangian}
\end{eqnarray}
The mass eigenstates of the up-type and down-type quarks 
are denoted by $u^a$ and $d^b$, with $a$ and $b$ being the generation indices.    
These quarks are also called  $(u,c,t,U)$ and $(d,s,b,D)$.    

     The decay width of $\BSgamma$ and the mixing parameter $x_s$ of 
$\BsBs$ depend on the $U$-quark mass  
$m_U$ and the CKM matrix elements $\VT$, $\VU$, $\ZBS$.  
The mass should be larger than the $t$-quark mass.  
The matrix elements are related to $V_{12}$, $V_{13}$, $V_{22}$, 
and $V_{23}$, which have been directly measured in experiments.  
From their experimental values~\cite{pdg}, a constraint 
\be
0.03<|\VT+\VU-\ZBS|<0.05  
\label{constraint1}
\ee
is derived.  The value of $\ZBS$ determines the magnitudes of 
flavor-changing interactions at the tree level.  
The upper bounds on the branching ratio of the decay 
$B\rightarrow K\mu^+\mu^-$~\cite{pdg} lead to a constraint 
\be
 |\ZBS| <2.0\times 10^{-3}.  
\label{constraint2}
\ee
In principle, the $U$-quark mass and the CKM matrix 
elements are related to each other through the mass 
matrices of the up-type and down-type quarks.  
However, these relations depend on many unknown factors for 
the mass matrices, so that various possibilities 
for the relations are not forbidden.  
Therefore, in our analyses, we take 
the model parameters $m_U$, $\VT$, $\VU$, and $\ZBS$ as 
independent of each other.  
For simplicity, these matrix elements are assumed to be real.  

\begin{table}
\begin{center}
\caption{The values of $\VU$ and $\ZBS$  
in Figs.~\ref{fig:bs} and~\ref{fig:bb}.}
\label{tab:ckm}
\vspace{0.2cm}
\begin{tabular}{c r r r r}
\hline
   & (i.a) & (i.b) & (ii.a) & (ii.b) \\
\hline
 $\VU$   & $-0.007$ & $-0.007$ & 0.006 & 0.006 \\
 $\ZBS$  & $-0.002$ & 0.002 & $-0.002$ & 0.002 \\
\hline
\end{tabular}
\end{center}
\end{table}
\vspace*{3pt}

\begin{figure}
\begin{center}
\begin{picture}(150,100)(0,0)
\ArrowLine(0,40)(45,40)
\Line(45,40)(105,40)
\ArrowLine(105,40)(150,40)
\PhotonArc(75,40)(30,0,180){4}{8.5}
\Photon(75,25)(85,0){4}{4}
\Text(-5,40)[]{$b$}
\Text(155,40)[]{$s$}
\Text(75,49)[]{$u\; c\; t\; U$}
\Text(55,80)[]{$W$}
\Text(95,15)[]{$\gamma\; g$}
\end{picture}
\end{center}
\caption{The diagram which gives a contribution to $C_7$ or 
$C_8$.   The photon or gluon line should be attached appropriately.}
\label{diagram:bs}
\end{figure}
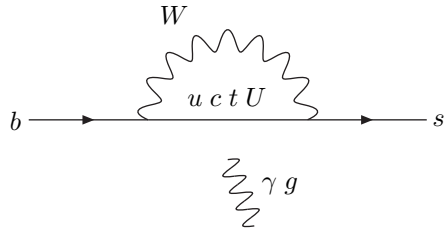

\section{Radiative Decay}

     The $B$-meson decay $\BSgamma$ is approximated by the  
$b$-quark decay $b\to s\gamma$.  
The relevant effective Hamiltonian is written as 
\be
{\cal H}_{eff} = -\frac{4G_F}{\r2} 
	\sum_{j=1}^8C_j(\mu) O_j(\mu),  
\ee
where $O_j$ and $C_j$ stand for respectively an operator of 
the $\Delta B = 1$ transition and a Wilson coefficient, 
with $\mu$ being the evaluated energy scale. 
Four-quark operators are denoted by $O_1-O_6$, 
while $O_7$ and $O_8$ represent the dipole operators for the photon 
and the gluon:  
\bea
O_7 &=& \frac{e}{16\pi^2}m_b\overline{s_L}\sigma^{\mu\nu}b_R F_{\mu\nu},  
                      \\
O_8 &=& \frac{g_s}{16\pi^2}m_b\overline{s_L}\sigma^{\mu\nu}T^ab_R 
             G_{\mu\nu}^a.  
\eea
Here $F_{\mu\nu}$ and $G_{\mu\nu}$ respectively stand for 
the electromagnetic and strong field strength tensors, 
with $T^a$ being the generator of SU(3).  
In the leading order approximation, the coefficients $C_7$ and $C_8$ 
receive contributions at $\mu=M_W$ from one-loop diagrams 
mediated by the $W$ boson, as shown in Fig.~\ref{diagram:bs}.  
Although the $Z$ and Higgs bosons mediate the decay through 
one-loop diagrams, their 
contributions can be neglected, owing mainly to the small 
magnitudes for the off-diagonal elements of $V^\dagger V$.  
The coefficient $C_7$ at $\mu = m_b$ is obtained by solving 
the renormalization group equations.  
We calculate the decay width, incorporating corrections to 
the next leading order for the anomalous 
dimensions and for the matrix element.  
The branching ratio for $\BSgamma$ is obtained by normalizing 
the decay width to that of the semi-leptonic decay 
$B \to X_ce\overline\nu$.  

\begin{figure}
\center
\psfig{file=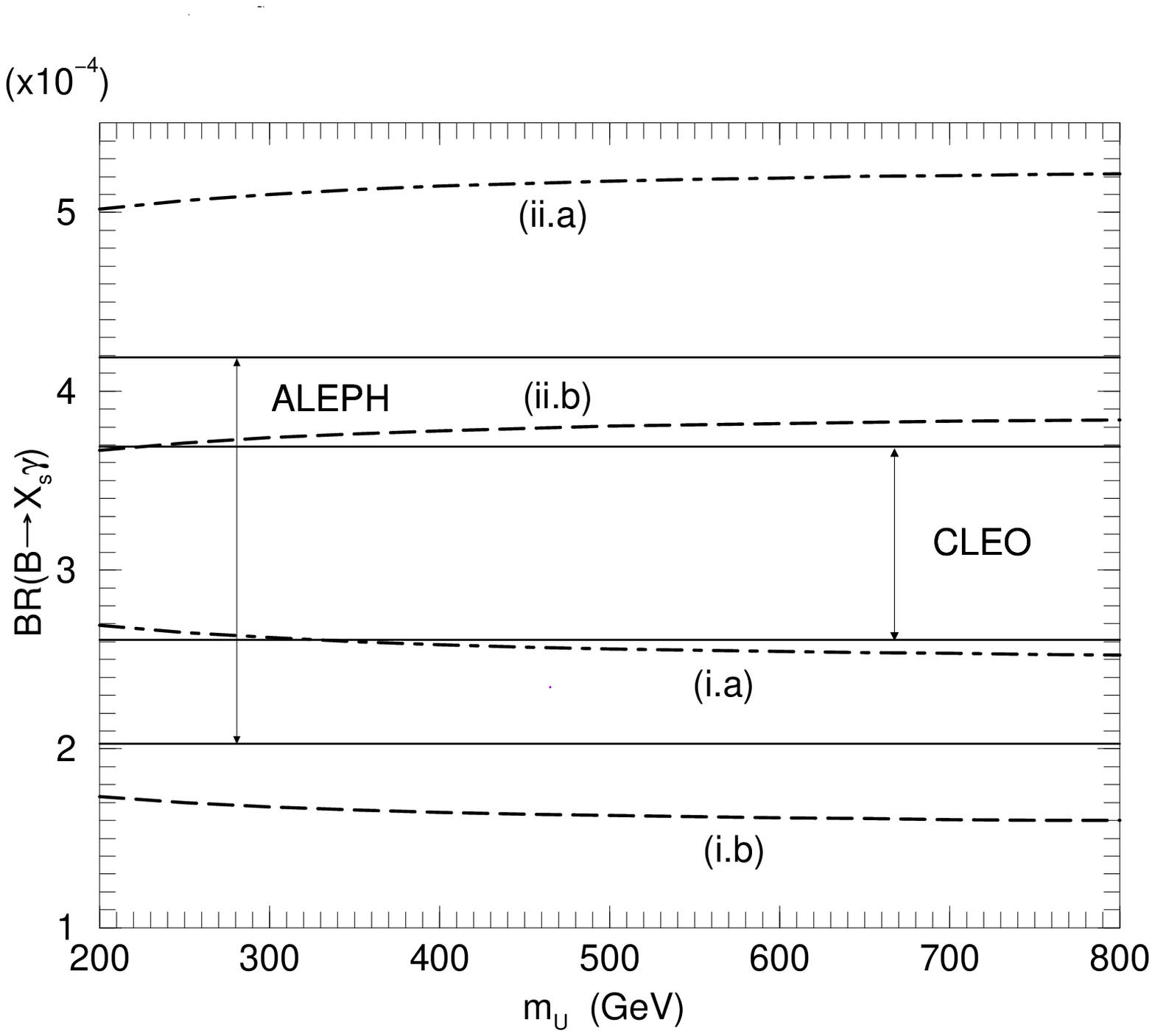,height=2.7in}
\caption{The branching ratio of $\BSgamma$ for $\VT=0.04$.}
\label{fig:bs}
\end{figure}

\begin{figure}
\center
\psfig{file=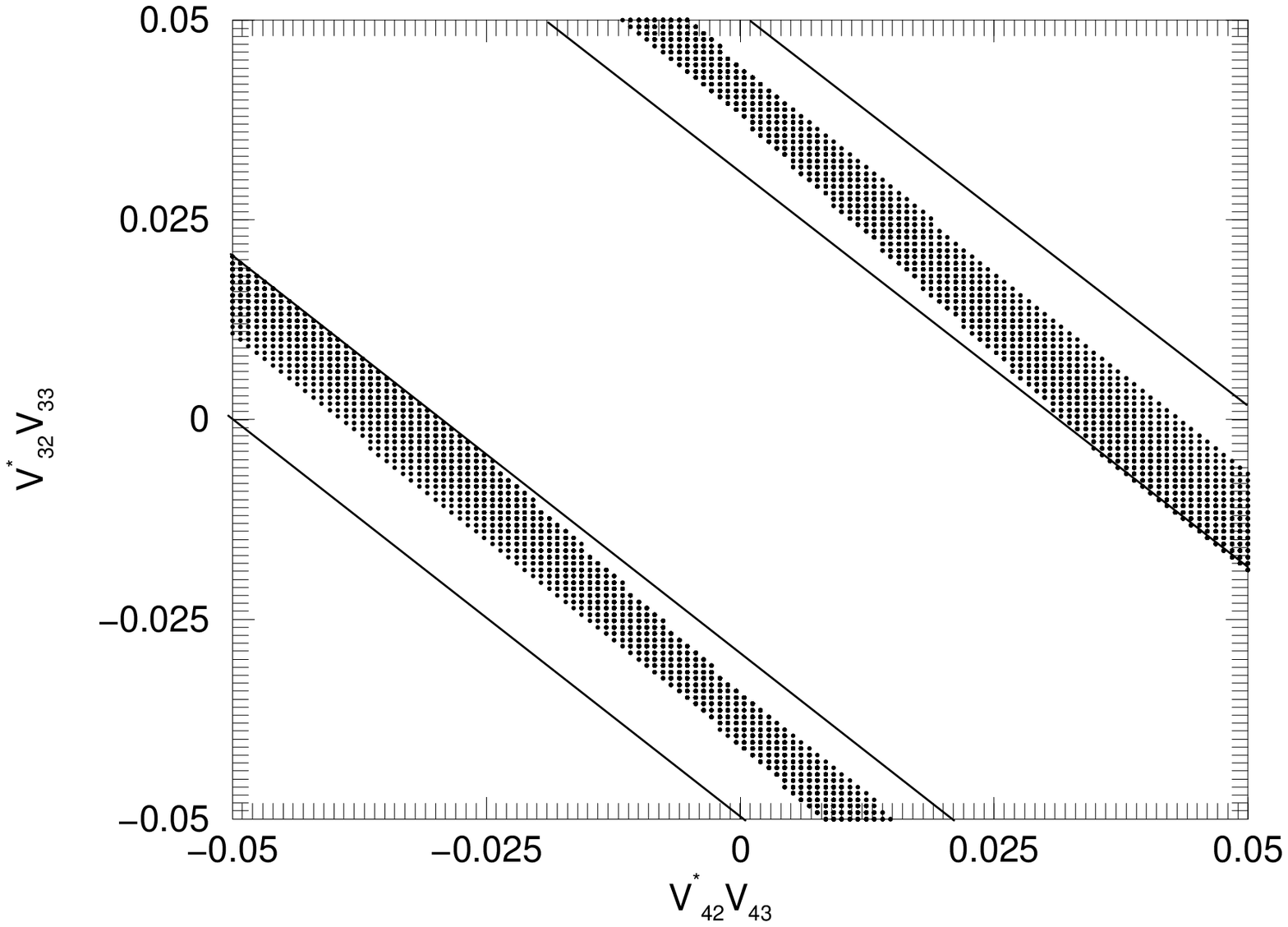,height=2.3in}
\caption{The allowed regions for $\VT$ and $\VU$ with
$\ZBS=1.0\times 10^{-3}$ and $m_U=200-800$ GeV.}
\label{fig:ckm}
\end{figure}

     The prediction for the branching ratio of $\BSgamma$ 
is shown in Fig.~\ref{fig:bs} as a function of the $U$-quark mass $m_U$ 
with $\VT=0.04$.  
Four curves (i.a)-(ii.b) correspond to the values of $\VU$ and $\ZBS$ 
listed in Table~\ref{tab:ckm}.  
The experimental bounds by CLEO~\cite{cleo} and ALEPH~\cite{aleph} 
are shown by solid lines.  
Sizable parts of the region which satisfies Eqs.~(\ref{constraint1}) 
and~(\ref{constraint2}) become inconsistent with the 
branching ratio of $\BSgamma$.  
The ranges $\VU<-0.007$ and $0.006<\VU$ are 
not allowed by the CLEO bounds, irrespectively of the value for $\ZBS$.  
The allowed ranges of $\VU$ and $\ZBS$ are not much dependent on $m_U$.   
For $|\VU/\VT|\gsim 0.1$, the branching ratio is non-trivially 
different from that of the SM, and  
any value within the experimental bounds could be predicted.  

     To see how the branching ratio of $\BSgamma$ 
constrains the CKM matrix elements, we show allowed regions 
for $\VT$ and $\VU$ in Fig.~\ref{fig:ckm}.  
The shaded regions are compatible with the experimental 
results for $\BSgamma$ by CLEO and  
for other CKM matrix elements translated as Eq.~(\ref{constraint1}).   
The regions between the solid lines satisfy the latter.  
We have taken the $U$-quark mass for 200 GeV $<m_U<$ 800 GeV 
and $\ZBS$ for $1\times 10^{-3}$.   
The branching ratio of $\BSgamma$ imposes stringent constrains 
on the CKM matrix elements in the VQM.  

\section{$\BsBs$ Mixing}

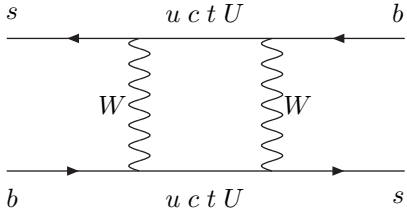
\begin{figure}
\begin{center}
\begin{picture}(150,100)
\ArrowLine(0,30)(50,30)
\Line(50,30)(100,30)
\ArrowLine(100,30)(150,30)
\ArrowLine(50,80)(0,80)
\Line(50,80)(100,80)
\ArrowLine(150,80)(100,80)
\Photon(50,30)(50,80){4}{6.5}
\Photon(100,30)(100,80){4}{6.5}
\Text(40,55)[]{$W$}
\Text(110,55)[]{$W$}
\Text(2,20)[]{$b$}
\Text(148,90)[]{$b$}
\Text(2,90)[]{$s$}
\Text(148,20)[]{$s$}
\Text(75,90)[]{$u\; c\; t\; U$}
\Text(75,20)[]{$u\; c\; t\; U$}
\end{picture}
\end{center}
\caption{The diagram which generates $\BsBs$ mixing.  
}
\label{diagram:bb}
\end{figure}

     The effective Hamiltonian for $\BsBs$ mixing is given by 
\be
{\cal H}_{eff} = \frac{G_F^2}{4\pi^2} M_W^2C(\mu) O(\mu),  
\ee
where $O$ denotes the four-quark operator of 
the $\Delta B = 2$ transition 
\be
O = \overline{s_L}\gamma_\mu b_L\overline{s_L}\gamma^\mu b_L, 
\ee
and $C$ stands for a coefficient.  
At $\mu=M_W$, the coefficient $C$ receives contributions from 
one-loop diagrams mediated by the $W$ boson, as shown 
in Fig. \ref{diagram:bb}. 
The mixing $\BsBs$ is also generated at the tree level by 
the $Z$ and Higgs bosons.  However, 
taking into account the experimental constraints on the 
flavor-changing interactions, the tree-level 
contributions turn out to be smaller than the one-loop 
contribution.  
Accordingly, the one-loop diagrams with the 
$Z$ or Higgs boson are negligible.  
The coefficient $C$ at $\mu=m_b$ receives QCD corrections.  

     One observable for $\BsBs$ mixing is the parameter $x_s$, 
which is defined by 
\be
     x_s=\frac{\Delta M_{B_s}}{\Gamma_{B_s}}.  
\ee
Here $\Delta M_{B_s}$ and $\Gamma_{B_s}$ represent the mass 
difference and the average width for the $B_s^0$-meson mass 
eigenstates.  The mass difference is induced dominantly by 
the short distance contributions.  
The decay constant and the bag factor of the $B_s^0$ meson 
for the matrix element are evaluated by lattice calculations.  

\begin{figure}
\center
\psfig{file=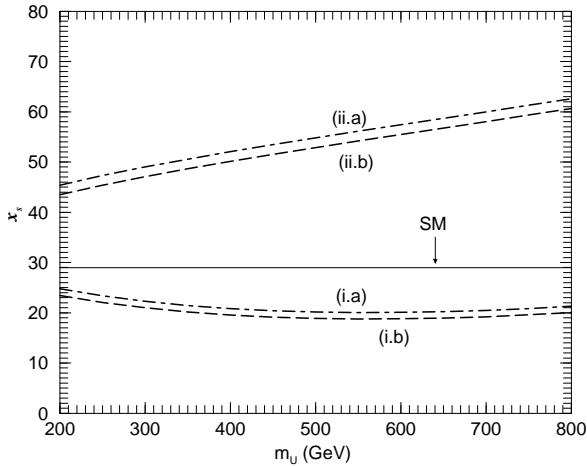,height=2.6in}
\caption{The mixing parameter $x_s$ for $\VT=0.04$.}
\label{fig:bb}
\end{figure}

     In Fig.~\ref{fig:bb} the mixing parameter $x_s$ 
is shown for the same parameter sets as those in Fig.~\ref{fig:bs}.  
The solid line represents the SM prediction.  
The value of $x_s$ varies in the region between the curves (i) and (ii) 
for the ranges of $\VU$ consistent with the branching ratio of $\BSgamma$.  
The mixing parameter can be larger than the SM value by 
as much as a factor of two.  
As the value of $\VT$ increases, the difference between 
the VQM and the SM becomes small.  
The parameter $x_s$ does not depend much on $\ZBS$, whereas 
a larger mass of the $U$ quark tends to give a larger value for $x_s$.  
For some ranges of the model parameters, the mixing parameter 
becomes smaller than the SM value, which does not happen in the 
supersymmetric standard model~\cite{branco}.  
 
     Within the framework of the SM, the measurement of $x_s$ 
determines the value of $\VT$.  
Then, the unitarity of the CKM matrix can be examined by summing 
$V^*_{12}V_{13}$, $V^*_{22}V_{23}$, and $\VT$.  
Another possible examination of the SM is to compare 
the value of $\VT$ thus determined with that measured by 
the $t$-quark decays.   
If a contradiction is found in these studies, 
the CKM matrix should be reanalyzed from the viewpoint of the VQM.  

\section{Discussions}

     We have discussed the effects of the VQM 
on the branching ratio for the radiative $B$-meson decay 
and the parameter $x_s$ for $\BsBs$ mixing.  
Although there are several new contributions, the $W$-mediated 
diagrams only yield sizable effects.
From the experimental results for the branching ratio, 
the values of $\VT$ and $\VU$ are severely constrained.  
These constraints do not strongly depend on the mass of the 
extra quark $U$.  
The VQM could make the branching ratio non-trivially different 
from the SM prediction.  
The mixing parameter may be either 
larger or smaller than the SM prediction by a factor of two or more.  

     In addition to causing effects on FCNC processes, the VQM 
induces $CP$ violation differently from the SM.  
The extended CKM matrix contains more than one physical complex phases, 
so that new $CP$-violating phenomena could occur.   
For instance, a $CP$-odd coupling is generated at the one-loop level in the 
gauge-boson self-interactions for $WWZ$~\cite{asakawa}, 
which could be examined in future experiments at linear $e^+e^-$ colliders.  

     The contributions of the VQM are 
also found in the semi-leptonic decay $B\to X_sl^+l^-$~\cite{ahmady}.  
This decay is induced both at the tree level and at the one-loop level.  
Its branching ratio and the forward-backward asymmetry for the 
final lepton could be different between the VQM and the SM.  
The decay rate asymmetry for $B$ and $\overline B$ is affected 
by the new source of $CP$ violation.  

\section*{Acknowledgments}

     This article is based on the works in collaboration with 
M.R. Ahmady, E. Asakawa, G.C. Cho, and A. Sugamoto.  
M.N. acknowledges a travel support from the conference.  
The work of M.A. is supported in part by the Grant-in-Aid for 
Scientific Research from the Ministry of Education, Science and
Culture, Japan.

\end{document}